\documentclass[twocolumn]{aastex701}
\usepackage{graphicx}	
\usepackage{amsmath}	

\usepackage{amssymb}	
\usepackage{bm}
\usepackage{amsfonts}
\usepackage{latexsym}
\usepackage{amsmath}
\usepackage{epsfig}
\usepackage{amsbsy}
\usepackage{enumitem}
\usepackage{multirow}
\usepackage[T1]{fontenc}

\begin{document}

\title{Thermal Deformations in Super-Eddington Magnetized Neutron Stars: Implications for Continuous Gravitational-Wave Detectability}

\correspondingauthor{Hong-Bo Li}
\email{lihb2020@pku.edu.cn}
\correspondingauthor{Yacheng Kang}
\email{yckang@stu.pku.edu.cn}
\correspondingauthor{Ren-Xin Xu}
\email{r.x.xu@pku.edu.cn}

\newcommand{\KIAA}{Kavli Institute for Astronomy and Astrophysics, 
           Peking University, Beijing 100871, People's Republic of China}
\newcommand{\PKUA}{Department of Astronomy, School of Physics, Peking University, Beijing 100871, People's Republic of China}
           
\author[0000-0002-4850-8351]{Hong-Bo Li}
\affiliation{\KIAA}
\email{lihb2020@pku.edu.cn} 

\author[0000-0001-7402-4927]{Yacheng Kang}
\affiliation{\KIAA}
\affiliation{\PKUA}
\email{yckang@stu.pku.edu.cn} 

\author[0000-0002-9042-3044]{Ren-Xin Xu}
\affiliation{\KIAA}
\affiliation{\PKUA}
\email{r.x.xu@pku.edu.cn} 

\begin{abstract}
Rapidly rotating neutron stars (NSs) are promising targets for continuous gravitational-wave (CGW) searches with current and next-generation ground-based GW detectors. In this work, we present the first study of thermal deformations in super-Eddington magnetized NSs with column accretion, where magnetic fields induce anisotropic heat conduction that leads to crustal temperature asymmetries. We compute the resulting mass quadrupole moments and estimate the associated CGW strain amplitudes. Our results show that Galactic magnetized NSs undergoing super-Eddington column accretion can emit detectable CGWs in upcoming observatories. Assuming a 2-yr coherent integration, the Einstein Telescope and Cosmic Explorer could detect such CGW signals from rapidly spinning NSs with spin periods $P \lesssim 20\,\rm ms$, while the LIGO O5 run may detect systems with $P \lesssim 6 \,{\rm ms}$. These findings suggest that super-Eddington magnetized NSs could represent a new class of CGW sources, providing a unique opportunity to probe the NS crust and bridge accretion physics with GW astronomy.

\end{abstract}


\section{Introduction}\label{sec: intro}

Ultraluminous X-ray sources (ULXs) are extragalactic objects with X-ray luminosities ($L_{\mathrm{X}}$) significantly exceeding ${10^{39}\,\mathrm{erg}\,\mathrm{s}^{-1}}$, and in some cases reaching up to ${10^{42}\,\mathrm{erg}\,\mathrm{s}^{-1}}$, well above the Eddington limit for a 10 solar mass black hole \citep{Kaaret:2017tcn, 2021AstBu..76....6F, King:2023nft}. These sources are generally found in active star-forming regions and are typically associated with young, luminous high-mass X-ray binaries (HMXBs) \citep{Roberts:2002ne, Gao:2003sr, Kovlakas:2020wfu}. Traditionally, ULXs have been interpreted as accreting stellar-mass \citep{Begelman:2006bi, Poutanen:2006uc} or intermediate-mass black holes \citep{Colbert:1999es,  2006ASPC..352..121M, Maccarone:2007dd, Koliopanos:2017aja}. However, this interpretation was challenged by the discovery of coherent X-ray pulsations from \mbox{M82 X-2} \citep{Bachetti:2014qsa}, which exhibits an X-ray luminosity of ${L_{\mathrm{X}} \simeq 1.8 \times 10^{40}\, \rm erg \, s^{-1}}$ and a pulse period of $1.37 \, \rm s$. Since black holes lack a solid surface, they are unlikely to generate such X-ray pulsations \citep{Kaaret:2017tcn}. Instead, the pulsating ULXs are naturally explained by accretion onto magnetized neutron stars (NSs), where infalling matter is channeled along magnetic field lines onto the poles, producing rotating hot spots and pulsed X-ray emission \citep{2010exru.book.....S}. These systems, known as ULX pulsars (ULXPs), extend the observed population of NS accretors reaching super-Eddington luminosities. The observed properties of currently known ULXPs are summarized in Table~\ref{tab: Table_1}.

\begin{table*}[ht]
    \renewcommand\arraystretch{1.5}
    \centering
    \caption{ Observed properties of known ULXPs. From left to right, the table lists: source name, period, period derivative, X-ray luminosity, distance, and the corresponding reference. The references are: $[{1}]$ \citet{Bachetti:2014qsa}; $[{2}]$ \citet{Israel:2016chx}; $[{3}]$ \citet{Vasilopoulos:2018vqh}; $[{4}]$ \citet{Sathyaprakash:2019stu}; $[{5}]$ \citet{Trudolyubov:2007nh}; $[{6}]$ \citet{Motch:2013zth}; $[{7}]$ \citet{Furst:2016mgk}; $[{8}]$ \citet{Israel:2016sxx}; $[{9}]$ \citet{Fuerst:2018ybc}; $[{10}]$ \citet{Quintin:2021beb}; $[{11}]$ \citet{Pintore:2021zbc}; $[{12}]$ \citet{Pintore:2025kcn}; $[{13}]$ \citet{NICER:2018evw}; $[{14}]$ \citet{Weng:2017qrc}; $[{15}]$ \citet{Vasilopoulos:2020jtu, Chandra:2020qlr}; $[{16}]$ \citet{Chandra:2020qlr}; $[{17}]$ \citet{RodriguezCastillo:2019esm}}
    \setlength{\tabcolsep}{0.75cm}
    {\begin{tabular}{l c c c c c c c c}
    \hline
    \hline
    Source   & $P$   & $\dot{P}$   & $L_{\rm X}$   & $d$   & Ref.
    \\[-0.5em]
             & $(\rm s) $   & $(10^{-9}\, \rm s \,s^{-1})$   & $(10^{39}\,\rm erg\, s^{-1})$   & $(\rm Mpc)$   & \\
    \hline
    M82 X-2                & 1.37   & $-0.2$   & $18$        & 3.6   & [1] \\
    NGC 5907 ULX-1         & 1.13   & $-3.8$   & $\sim 100$  & 17.1  & [2] \\
    NGC 300  ULX-1         & 16.6   & $+150$   & $4.7$       & 2     & [3] \\
    NGC 1313 X-2           & 1.5    & $-13.8$  & $\sim 20$   & 4     & [4] \\
    NGC 2403 ULX           & 18     & $-100$   & $1.2$       & 3.2   & [5] \\   
    NGC 7793 P13           & 0.42   & $-0.03$  & $\sim 10$   & 3.6   & [6-9] \\
    NGC 7793 ULX-4         & 0.4    & $-35$    & $3.4$       &3.9    & [10] \\
    NGC 4559 X-7           & 2.6    & $+1$     & $20$        & 7.5   & [11-12] \\
    Swift J0243.6$+$6124   & 9.86   & $-22$    & $2$         & 0.007 & [13] \\
    SMC X-3                & 7.8    & $-6.46$  & $1.2$       & 0.062 & [14] \\
    RX J0209.6$-$7427      & 9.29   & $-17.5$  & $1.6$       & 0.06  & [15-16]\\
    M51 ULX-7              & 2.8    & $-0.15$  & $4$         & 8.6   & [17]\\
    \hline
    \end{tabular}}
    \label{tab: Table_1}
\end{table*}

Notably, recent population synthesis studies further suggest that NSs in low- and intermediate-mass X-ray binaries (LMXBs and IMXBs) can reach ULX luminosities during intense mass-transfer phases \citep{King:2001pm, Shao:2015eha, Wiktorowicz:2018duf, Quast:2019csj, Misra:2020oox}. However, the detailed physical origin of the extreme luminosities observed in ULXPs are still under debate. Proposed models span a wide range of NS surface magnetic field strengths, from $10^{11}\, \rm G$, where the luminosity may be boosted by strong outflows and geometrical beaming \citep{Middleton:2016qss, 2017MNRAS.471L..71M, King:2019jvg, Lasota:2023xef, Kovlakas:2025qig}, to magnetar-like fields of $10^{14}\, \rm G $, which can suppress radiation pressure \citep{1976MNRAS.175..395B, Tong:2014esa, DallOsso:2015nzq, Eksi:2014lya, Kovlakas:2025qig} and confine accretion columns above the magnetic poles \citep{Mushtukov:2015zea, 2020PASJ...72...12I}. In addition to magnetic field strength, several other factors may also significantly affect accretion dynamics in ULXPs, including non-dipolar magnetic field configurations \citep{Israel:2016chx, Tsygankov:2018bcz}, enhanced neutrino cooling \citep{Mushtukov:2018pvq, 2019MNRAS.485L.131M, Asthana:2023vvk, Mushtukov:2024efo, 2025arXiv250913732L}, and photon bubble instabilities \citep{1992ApJ...388..561A, Begelman:2006ap, 2021MNRAS.508..617Z}. In view of this, electromagnetic observations alone may not allow us to fully infer the intrinsic properties of these super-Eddington magnetized NSs.

On the other hand, rapidly rotating NSs can emit continuous gravitational wave (CGW) via an oscillating mass quadrupole moment, arising from various possible mechanisms such as elastic strain \citep{Haskell:2006sv, Johnson-McDaniel:2012wbj, Gittins:2020cvx, Morales:2022wxs, Morales:2024hwl}, magnetic deformation \citep{Melatos:2005ez, Haskell:2007bh, Fujisawa:2022dzp}, and $r$-mode instabilities \citep{Andersson:1998qs, Levin:1998wa, Arras:2002dw, Bondarescu:2008qx}. For LMXBs, \citet{Singh:2019dgy} also suggested that magnetic stresses can induce displacements of electron-capture layers in the accreted crust, potentially giving rise to asymmetries.The self-consistent equilibrium structure of the polar magnetic mountains has been extensively studied under the Newtonian gravity \citep{Payne:2004vt, Payne:2007kn, Vigelius:2008fv, Wette:2009gj, Vigelius:2009yn, Vigelius:2009eg, Priymak:2011zv, Suvorov:2018lvu, Suvorov:2020jqf, Fujisawa:2022dzp}. More recent studies extend these investigations into the general relativistic regime \citep{Rossetto:2023btr, Rossetto:2024oes}. Motivated by this, super-Eddington magnetized NSs may also represent promising CGW targets. 
In particular, magnetically channeled column accretion from the companion can induce non-axisymmetric crustal deformations in rapidly spinning NSs, generating significant CGW signals. Even in the absence of a detection, stringent upper limits on CGW amplitudes from these super-Eddington magnetized NSs can provide meaningful constraints on their accretion dynamics and internal structure.

In this work, we present the first quantitative study of thermal deformations in super-Eddington magnetized NSs with column accretion, focusing on crustal temperature asymmetries induced by magnetically modulated anisotropic heat conduction. We compute the resulting mass quadrupole moments and evaluate the detectability of the associated CGW emission with current and next-generation ground-based GW detectors, including the Laser Interferometer Gravitational-wave Observatory \cite[LIGO;][]{KAGRA:2013rdx}, the Einstein Telescope \cite[ET;][]{Branchesi:2023mws}, and the Cosmic Explorer \cite[CE;][]{Srivastava:2022slt}. Our results indicate that Galactic magnetized NSs undergoing super-Eddington column accretion can produce detectable CGWs in upcoming observatories, representing a new class of CGW sources. Assuming a 2-yr coherent integration, the ET and CE could detect such CGW signals from rapidly spinning NSs with spin periods $P \lesssim 20\,\rm ms$, while the LIGO O5 run may detect systems with $P \lesssim 6 \,{\rm ms}$. These findings suggest that super-Eddington magnetized NSs provide a unique opportunity to probe the NS crust and bridge accretion physics with GW astronomy.

The structure of this paper is organized as follows. Section~\ref{sec: column} describes the construction of the accretion column model above these super-Eddington magnetized NSs. In Section~\ref{sec: Temperature perturbation}, we calculate their thermal crustal deformations. Section~\ref{sec: detections} discusses the detectability of the associated CGW signals, and Section~\ref{sec: conclusion} presents our conclusions.

\section{Structure of the accretion column}
\label{sec: column}

Super-Eddington accretion onto magnetized NSs generates radiation pressure-dominated shocks above the NS surface, leading to the formation of accretion columns \citep{1975PASJ...27..311I, Takahashi:2017nxo, 2020PASJ...72...12I, Inoue:2020aut, 2020PASJ...72...15K, Inoue:2023cph}. The structure and temperature of these columns are primarily determined by the accretion rate and the optical depth of the inflowing material. The governing dynamical equations can be expressed as \citep{1975PASJ...27..311I, 2020PASJ...72...12I}
\begin{align} 
\frac{{\rm d}\varepsilon}{{\rm d}r} & = \frac{3}{4} ( \frac{\varepsilon}{r_{\rm D}} - \frac{GM}{r^{2}} )  \label{eq: eq_1} \,, \\ 
\frac{{\rm d}\rho}{{\rm d}r} & = -\frac{3\rho}{4\varepsilon} ( \frac{\varepsilon}{r_{\rm D}} + 3  \frac{GM}{r^{2}} )  
\label{eq: eq_2}\,, 
\end{align}
where $\varepsilon$ and $\rho$ denote the radiation energy per unit mass and the mass density within the accretion column, respectively; $M$ is the NS mass; and the photon diffusion length $r_{\rm D}$ is given by $r_{\rm D} =1.4 \times 10^{5} \left(\frac{\dot{M}}{10^{17} \rm{g \; s}^{-1}} \right) \; \rm{cm}$, where $\dot{M}$ is the mass accretion rate onto the NS. 

\begin{figure*}[ht]
    \centering
    \includegraphics[width=18cm]{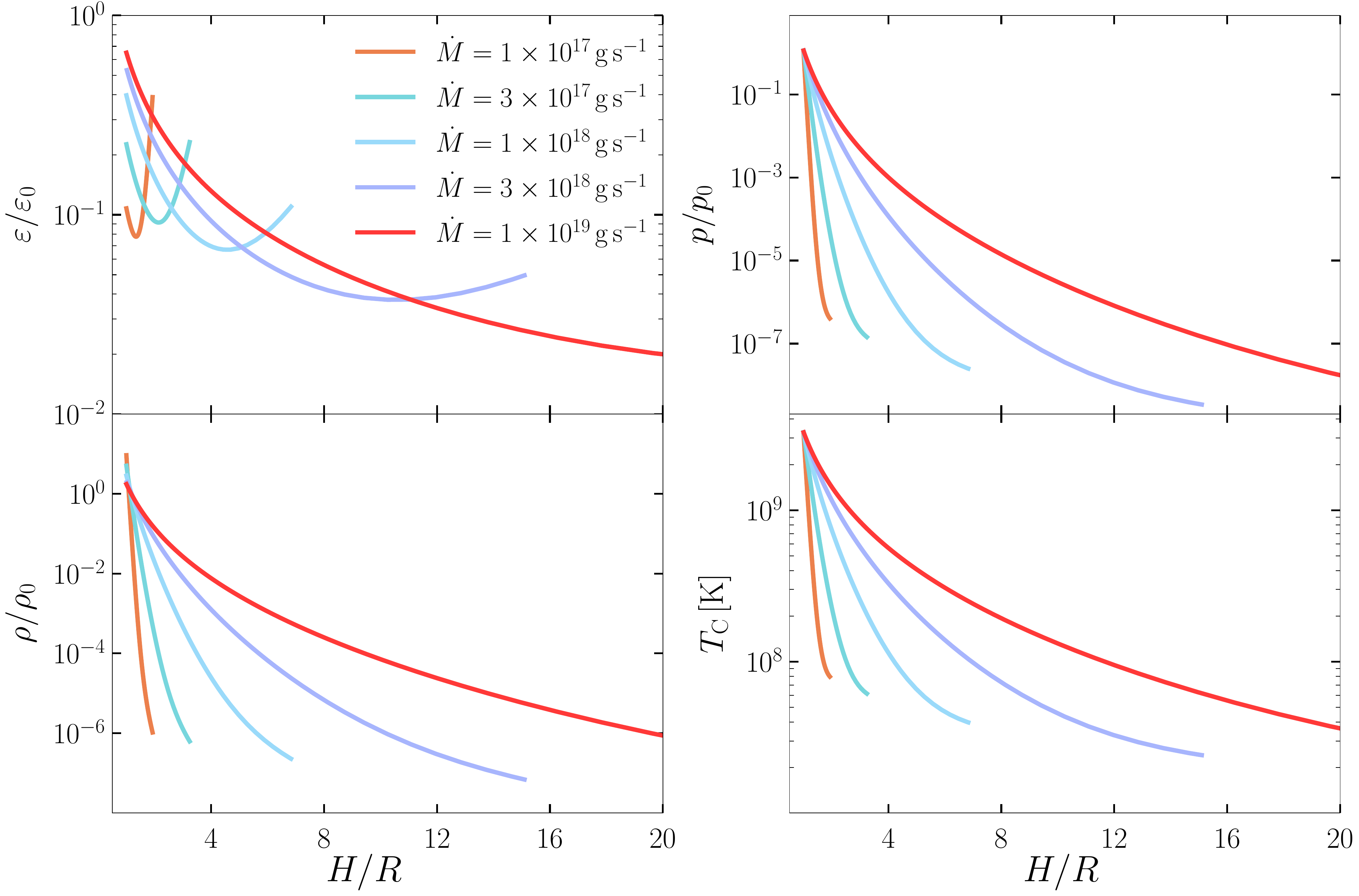}
    \caption{Profiles of physical quantities within the accretion column above super-Eddington magnetized NSs as functions of height. Different colors indicate different mass accretion rates. Note that we adopt a typical NS model with $M = 1.4 \, M_{\odot}$ and $R = 10^{6}\, \rm cm$. The left panels present the radiation energy density and mass density, while the right panels show the corresponding pressure and temperature profiles. The characteristic temperature, $T_{\rm C}$, is calculated via $T_{\rm C} = (\rho \varepsilon/a)^{1/4}$, where $a$ is the radiation constant. Assuming a surface magnetic field strength of $B = 10^{12}\, \rm G $, we define the normalization parameters as $\varepsilon_{0} = GM/R$, $\rho_{0}=3 B^2 R/ (8 \pi G M)$, and $p_0 =B^2/ (8\pi) $.}
    \label{fig: accretion_column}
\end{figure*}

Before quantitatively analyzing thermal deformations in super-Eddington magnetized NSs, we should first solve Equations~(\ref{eq: eq_1}) and (\ref{eq: eq_2}) to determine the physical properties of the accretion columns. Taking a NS model with $M = 1.4 \, M_{\odot}$ and $R = 10^{6}\, \rm cm$ as an example, we present in Figure~\ref{fig: accretion_column} the radiation energy, mass density, pressure, and temperature of accretion columns as functions of the column height for different mass accretion rates. Details of the numerical treatments are provided in Appendix~\ref{sec: Appendix_I}. One can find that an extremely high mass accretion rate results in a taller accretion column. Moreover, the temperature at the base of the column remains largely insensitive to the accretion rate, as the radiation pressure is counterbalanced by the magnetic pressure. The characteristic temperature at the column base is approximately $4 \times 10^9 \, \rm K$, which exceeds that typically observed in most known LMXB systems \citep{1999ApJ...524.1014S}.

\section{Temperature perturbation and crust deformation}\label{sec: Temperature perturbation}

\begin{figure*}
    \centering
    \includegraphics[width=14cm]{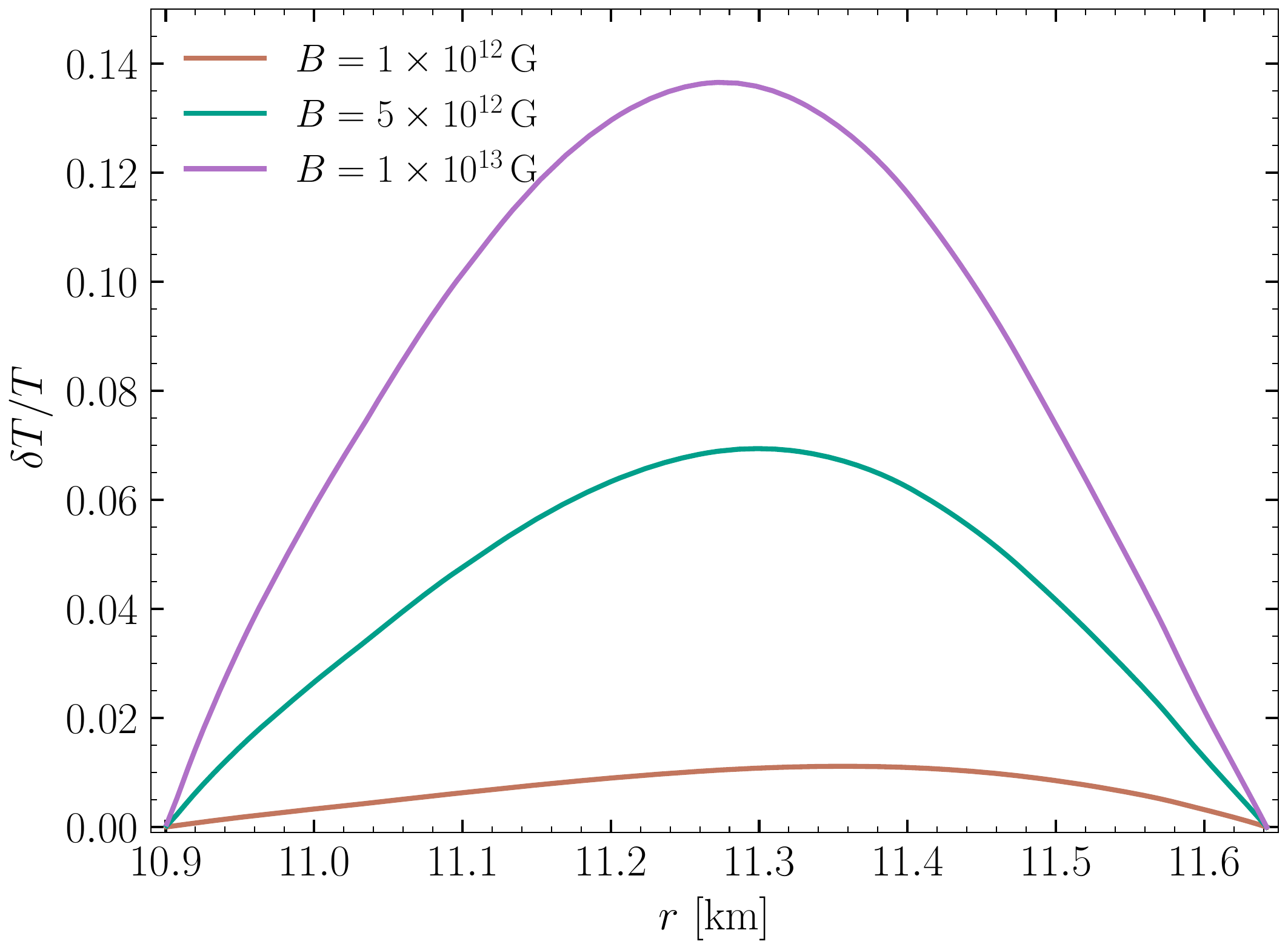}
    \caption{Temperature perturbations in the crust of ULXPs for different magnetic field strengths, assuming a fixed accretion rate of $\dot{M} = 5 \times 10^{18}\, \rm g\, s^{-1}$.}
    \label{fig: Delt_Temperature}
\end{figure*}

We now investigate the thermal structure of magnetized NSs undergoing super-Eddington column accretion and derive the resulting temperature perturbations. Assuming that the magnetized NS is spherically symmetric and in a steady-state thermal equilibrium, the heat flux driven by column accretion obeys Fourier's law as
\begin{equation} 
{\bf F} = - \kappa \nabla T \,, 
\end{equation}
where $\kappa$ denotes the thermal conductivity. 
In Appendix~\ref{sec: Appendix_II}, we provide a detailed discussion of the equations used to determine the thermal structure of an accreting NS, including the accretion heating, neutrino cooling, thermal conductivity of the NS crust, and relevant boundary conditions.
Importantly, the boundary condition is set by the characteristic temperature at the base of the accretion column. The presence of internal toroidal magnetic fields in super-Eddington magnetized NSs can introduce anisotropy in the thermal conductivity, thereby generating significant temperature gradients across the crust \citep{Geppert:2004sd, Geppert:2005dm, Page:2007br, Aguilera:2007xk}. 

Following \citet{Osborne:2019iph}, we compute the temperature perturbations in these super-Eddington magnetized NSs with column accretion. This approach is valid for magnetic field strengths $B \lesssim 10^{13}\, \rm G$ \citep{Osborne:2019iph, Hutchins:2022chj}. Note that in this section, we adopt the DH NS equation of state from  \citet{Douchin:2001sv} for calculating temperature perturbations, which yields a NS mass of $M \simeq 1.84 \, M_{\odot}$ and radius $R \simeq 1.17 \times 10^{6}\, \rm cm$. These values differ from the typical NS model adopted in Section~\ref{sec: column}. Further details on the calculations of magnetic-field-induced perturbations in the thermal structure are provided in Appendix~\ref{sec: Appendix_III}. Our results, shown in Figure~\ref{fig: Delt_Temperature}, illustrate the dependence of the relative temperature perturbation $\delta T / T$ on magnetic field strength. An increase in magnetic field strength leads to more prominent temperature asymmetries within the crust. Compared with typical values in most LMXBs \citep[see e.g.,][]{Osborne:2019iph}, the temperature perturbations in super-Eddington magnetized NSs can be enhanced by up to four orders of magnitude. This significant amplification arises from the combination of a higher base temperature at the accretion column foot and the characteristically stronger internal magnetic fields.

As noted by \citet{Ushomirsky:2000ax}, electron capture rates in the NS crust are highly sensitive to temperature, particularly when it exceeds $2 \times 10^8\, \rm K$. Elevated temperatures can enhance electron capture transitions relative to cooler regions, thereby inducing compositional and structural asymmetries. Consequently, large-scale temperature perturbations in magnetized NSs undergoing super-Eddington column accretion can induce elastic deformations in the crust. When such asymmetries are misaligned with the NS rotation axis, they lead to non-axisymmetric density distributions and a resulting mass quadrupole moment $Q_{22}$. In this work, we apply the self-consistent numerical approach developed by \citet{Li:2024dvx} to compute $Q_{22}$. A detailed discussion of the methodology is presented in Appendix~\ref{sec: Appendix_IV}. We find that the corresponding values of $Q_{22}$ are $4.5 \times 10^{37}\, \rm g \, cm^2$ for $B = 10^{12}\, \rm G$, $8.6 \times 10^{37}\, \rm g \, cm^2$ for $B =  5 \times 10^{12}\, \rm G$, and $1.2 \times 10^{38}\, \rm g \, cm^2$ for $B = 10^{13}\, \rm G$, respectively.

\begin{figure*}[ht]
    \centering
    \includegraphics[width=15cm]{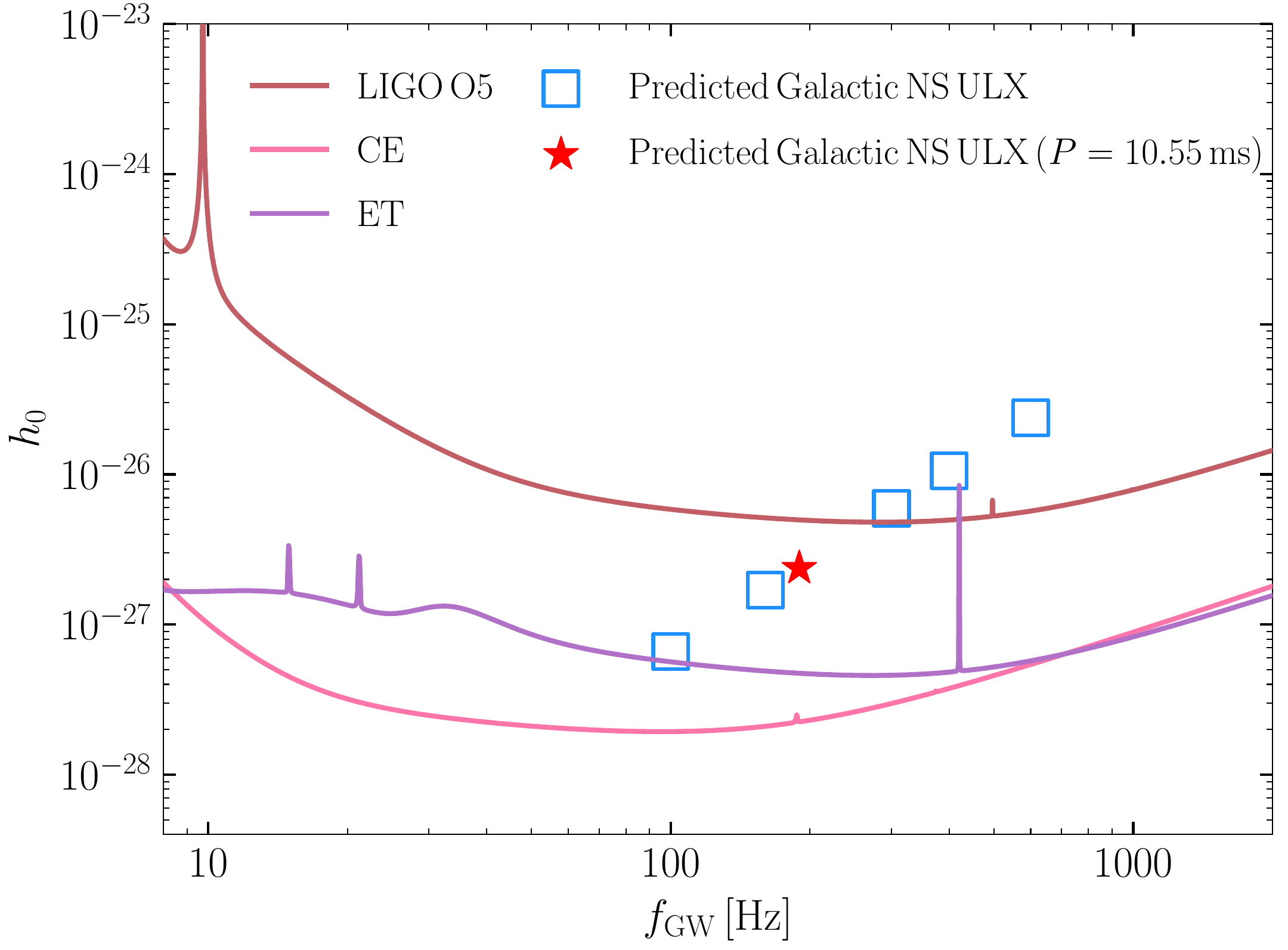}
    \caption{GW strain amplitude as a function of the frequency. Open blue squares represent predicted Galactic NS ULXs with spin periods of ${P = 20 \,{\rm ms}}$, ${P = 12.5 \,{\rm ms}}$, ${P =6\,{\rm ms}}$, ${P =5\,{\rm ms}}$, and ${P =3\,{\rm ms}}$. The filled red star marks a predicted Galactic NS ULX system with PSR J1928$+$1815-like spin period of ${P = 10.55 \,{\rm ms}}$. All these predicted NS ULXs are assumed to be located at a distance of $d = 8\, {\rm kpc}$. Sensitivity curves for LIGO O5, ET, and CE assume a coherent integration time of two years \citep[see e.g.,][]{Watts:2008qw}.}
    \label{fig: strain}
\end{figure*}

\section{CGW detectability}
\label{sec: detections}

Rapidly rotating, super-Eddington magnetized NSs with asymmetric deformations can generate CGWs via quadrupole radiation \citep{1979PhRvD..20..351Z, Jaranowski:1998qm}. To assess the detectability of such signals, we express our results in terms of the stellar ellipticity, defined as $ \epsilon \equiv \sqrt{\frac{8 \pi}{15}} \frac{Q_{22}}{I_{\rm zz}}$, where $I_{\rm zz} \simeq 10^{45}\, \rm g\, cm^2$ is the principal moment of inertia. The corresponding GW strain amplitude $h_{0}$ is given by \citep{LIGOScientific:2019yhl, Reed:2021scb}
\begin{equation}\label{eq: strain_2}
   h_{0} = \frac{4 \pi^2 G}{c^4}\frac{I_{\rm zz} \epsilon \, f^2_{\rm GW}}{d} \,,
\end{equation}
where $f_{\rm GW} = 2 \nu$ is the GW frequency corresponding to a magnetized NS with spin frequency $\nu$, and $d$ is the source distance. In this work, we suggest that the ellipticity resulting from the temperature-induced asymmetry can be approximately expressed as $\epsilon  \simeq  2 \times 10^{-7}\, B_{12} $, where $B_{12}$ is the surface magnetic field strength in units of $10^{12}\, \rm G$. 
In super-Eddington magnetized neutron stars with column accretion, the magnetic field channels the accretion flow onto the magnetic poles and can introduce anisotropic thermal conductivity in the crust, particularly in the presence of internal toroidal magnetic fields \citep{Geppert:2004sd, Geppert:2005dm, Page:2007br, Aguilera:2007xk}. This anisotropic heat transport leads to non-axisymmetric temperature distributions, producing crustal density perturbations and a corresponding mass quadrupole moment. Our numerical calculations of the temperature perturbations and the resulting quadrupole moments indicate that the corresponding stellar ellipticity typically lies in the range $\epsilon \sim 10^{-8}$ -- $10^{-7}$ for magnetic fields $B \simeq 10^{12}$ -- $10^{13} \, \rm G$. These results can be approximately summarized by the scaling relation $\epsilon  \sim (1 - 2 ) \times 10^{-7}\, B_{12} $. This expression should be regarded as an order-of-magnitude estimate inferred from the numerical calculations rather than a strict analytical relation. We employ Equation (4) to assess the detectability of CGWs from all known ULXPs listed in Table 1.


For comparison, the ellipticity produced by purely magnetic stresses inside NSs is typically expected to scale as $\epsilon \propto B^2$ for stars with normal (non-superconducting) cores. In this class of models, explored for example by \cite{2002PhRvD..66h4025C, 2008MNRAS.385..531H, 2009MNRAS.395.2162L, 2009MNRAS.394.1711H}, the deformation arises from the direct magnetic distortion of the stellar structure and can be understood, at the order-of-magnitude level, as being set by the ratio of the magnetostatic energy to the gravitational binding energy. By contrast, the ellipticity obtained in our work is several orders of magnitude larger because the deformation considered here is not generated directly by internal magnetic stresses, but instead arises from magnetically regulated thermal asymmetries associated with super-Eddington accretion.

In this respect, our mechanism is more closely related to the thermal-mountain scenario discussed by \cite{Osborne:2019iph, Hutchins:2022chj}, in which the magnetic field acts primarily by inducing temperature non-axisymmetry through anisotropic thermal conductivity, rather than by directly deforming the star.  The approximately linear scaling adopted here therefore reflects a thermally mediated deformation mechanism, $\epsilon  \simeq B$, within the perturbative regime explored by our calculations. The above scaling is derived under physical conditions characteristic of ULX pulsars, namely high accretion luminosities $L_{\rm X} \gtrsim 10^{39} \,\rm erg \,s^{-1}$ and magnetic fields $B \sim 10^{11}$ -- $10^{13} \,\rm G$.  In systems with significantly lower magnetic fields and accretion luminosities, such as typical LMXBs, magnetic confinement of the accretion flow and the associated anisotropic thermal transport are expected to be weaker, leading to smaller temperature asymmetries and therefore smaller stellar deformations.

Unfortunately, we find that the expected GW strain amplitudes $h_{0}$ from these sources lie below the sensitivity of LIGO's O5 run \citep{KAGRA:2013rdx} for a 2-yr coherent integration \citep{Watts:2008qw}, and are far below the detection capabilities of ET and CE. This non-detectability arises primarily from two factors. First, except for Swift J0243.6$+$6124, all detected ULXPs are extragalactic, leading to substantial GW signal attenuation due to their large distances. Second, the spin frequencies of these ULXPs are insufficiently high to place the resulting GW signals within the peak sensitivity range of current and next-generation detectors.

Although the results above may seem discouraging given the current detected ULXPs, recent population synthesis studies suggest that there could be $\sim 7-20$ NS ULXs with helium-star donors in our Galaxy \citep{Shao:2015eha, 2019ApJ...886..118S, 2020A&A...642A.174M, Gao:2022fsg, Misra:2023oqk, Li:2024njw}. Notably, a Galactic detached pulsar + helium star binary, PSR J1928$+$1815, with the NS spin period of $P = 10.55 \,{\rm ms}$, has recently been discovered by the Five-hundred-meter Aperture Spherical radio Telescope \citep[FAST;][]{Yang:2025khi}. Such nearby NS + helium star binaries in our Galaxy may represent an early stage preceding intense mass transfer. Once they enter the super-Eddington accretion phases, they could significantly contribute to both the Galactic ULX population and the detectable CGW signal. 

More importantly, these Galactic NS ULXs can be spun up (recycled) during their accretion phases, ultimately evolving into millisecond pulsar $+$ white dwarf systems, which are observed as intermediate-mass binary pulsars (IMBPs). Consequently, the observed spin periods of IMBPs can serve as empirical upper bounds on the spin frequencies of magnetized NSs in their earlier super-Eddington accretion phases. We find that several known IMBPs exhibit NS spin periods shorter than $P \simeq 20 \,{\rm ms}$; for instance, PSR J1614$-$2230 has an exceptionally short spin period of $P \simeq 3 \,{\rm ms}$ \citep{2011MNRAS.416.2130T, Tauris:2012jp, Li:2024njw}, supporting the potential existence of Galactic NS ULXs with spin periods below $P = 20 \,{\rm ms}$. In view of this, Figure~\ref{fig: strain} presents the predicted CGW strain amplitudes for Galactic NS ULXs with various fixed spin periods and their corresponding detectabilities. Assuming a 2-yr coherent integration time, we find that the predicted Galactic NS ULXs with spin periods shorter than $20 \, \rm ms$ could be detected by next-generation ground-based GW observatories, including the ET and CE. The LIGO O5 run may detect Galactic NS ULXs with $P \lesssim 6 \,{\rm ms}$. Although no such rapidly spinning NS ULXs have been observed in our Galaxy to date, our results highlight their potential as promising CGW sources, should a Galactic NS ULX system with $P \lesssim 20 \,{\rm ms}$ indeed exist. We encourage future population studies to further explore the occurrence and properties of these fast-spinning super-Eddington magnetized NSs.

\section{Conclusion}
\label{sec: conclusion}

In this work, we present the first study of thermal deformations in super-Eddington magnetized NSs undergoing column accretion and investigate their implications for CGW emission. Focusing on radiation-pressure-dominated accretion columns, we compute their base temperatures and demonstrate how magnetically induced thermal perturbations produce localized asymmetries. These asymmetries can generate a finite mass quadrupole moment, potentially leading to observable CGW signals. Given the strong dependence of detectability on the spin period, our results indicate that next-generation ground-based GW detectors, such as the ET and CE, could observe such CGW signals from Galactic NS ULXs with spin periods $P \lesssim 20\,\rm ms$, while the upcoming LIGO O5 run can only detect the fast-spinning super-Eddington magnetized NSs with ${P \lesssim 6 \,{\rm ms}}$.

Future work incorporating more realistic models of internal magnetic field configurations \citep{2004A&A...426..267G, 2006A&A...457..937G, 2007A&A...470..303P, 2025arXiv250906699P} and updated microphysics of the accreted crust \citep{2018A&A...620A.105F, 2022A&A...665A..74F} will enable more accurate predictions of thermal deformation and CGW amplitudes. Such refinements will not only improve detectability estimates for future GW detectors but also help optimize search strategies, offering deeper insights into the internal structure of magnetized NSs. Further dedicated studies in this direction are encouraged.

\begin{acknowledgments}

We thank Lijing Shao, Wencong Chen, Shijie Gao and Yiming Hu for helpful discussions. This work was supported by the National Natural Science Foundation of China (12447148), the China Postdoctoral Science Foundation (2024M760081), the National SKA Program of China (2020SKA0120100), and the High-Performance Computing Platform of Peking University. YK is supported by the China Scholarship Council (CSC).

\end{acknowledgments}

\newpage

\appendix

\section{Numerical treatment of accretion column structure}
\label{sec: Appendix_I}
\setcounter{equation}{0}
 \setcounter{figure}{0}
 \setcounter{table}{0}
\renewcommand{\thesection}{\Alph{section}}
\renewcommand{\theequation}{A\arabic{equation}}
 \renewcommand{\thesection}{\Alph{section}}
 \renewcommand{\thesubsection}{\thesection.\arabic{subsection}}
 \renewcommand{\thesubsubsection}{\thesubsection.\arabic{subsubsection}}
 \makeatother
 

Assuming one-dimensional, steady-state super-Eddington column accretion, the structure of the accretion column is determined by the conservation of mass, momentum, and energy \citep[see e.g.,][]{1975PASJ...27..311I, 2020PASJ...72...12I}. Equation~(\ref{eq: eq_1}) describes the evolution of the energy density, balancing the gravitational acceleration $(GM/r^{2})$ against radiative dissipation; while Equation~(\ref{eq: eq_2}) governs the density stratification, where the compressive effects of gravity are balanced by radiative pressure. Together, the two equations jointly define the vertical structure of the accretion column and the associated energy transport in the polar regions.

At the top of the polar accretion columns, the mass density and radiation energy density are set as follows \citep{1975PASJ...27..311I, 2020PASJ...72...12I}
\begin{equation}\label{eq: boundary conditions_1}
\begin{aligned} 
  \rho_{\rm top} &= 7 \rho_{\rm F}(r_{\rm S}) \,,\\
  \varepsilon_{\rm top} &= \frac{3}{4}\frac{GM}{r_{\rm S}} \,.
\end{aligned} 
\end{equation}
Here, $\rho_{\rm F}$ denotes the matter density in the free-fall region at the top boundary location $r_{\rm S}$. The position $r_{\rm S}$ is iteratively adjusted to ensure consistency with the bottom boundary condition. Specifically, we numerically integrate the dynamical equations inward from an initial guess for $r_{\rm S}$ toward the stellar surface. At the surface, the bottom boundary is defined by the balance between gas pressure and magnetic pressure:
\begin{equation}\label{eq: boundary conditions_3}
  p_{*} = \frac{\rho_{*} \varepsilon_{*}}{3} =\frac{ B^2_{*}}{8 \pi}\,,
\end{equation}
where the subscript ``$*$'' denotes quantities evaluated at the stellar surface.

\section{Thermal structure of an accreting NS}
\label{sec: Appendix_II}
\setcounter{equation}{0}
\renewcommand{\theequation}{B\arabic{equation}}

In this Appendix, we calculate a steady-state thermal structure based on the Newtonian formulation \citep{Ushomirsky:2000ax, Li:2024dvx}.  The temperature profile is spherically symmetric and independent of the time.  From energy conservation, the heat flux $\bf F$ is related to the net rate of heat energy generation per unit volume $ Q$ via
\begin{equation} 
\nabla \cdot {\bf F} = Q = Q_{\rm nuc} - Q_{\rm neu} \,,
\end{equation}
where $Q_{\rm nuc}$ is the local energy deposited via nuclear reactions, and $Q_{\rm neu}$ is the local energy loss due to neutrino emission. $\bf F$ is related to the temperature $T$ by the Fourier's law,
\begin{equation} 
{\bf F} = - \kappa \nabla T \,, 
\end{equation}
where $\kappa$ is the thermal conductivity.  Assuming the background structure is spherically symmetric, $\bf F$ only depends on the radial coordinate $r$. Therefore, the following differential equations for the heat flux $F$ and temperature $T$ can be written as
\begin{align}
 \frac{{\rm d} F}{{\rm d} r} & = Q - \frac{2}{r} F  
 \label{eq: dFdr} \,, \\ 
\frac{{\rm d} T}{{\rm d} r} & = - \frac{1}{\kappa}  F   \label{eq: dTdr}\,.  
\end{align}
To solve the above equations, we will consider the heating term $Q_{\rm nuc}$, neutrino cooling $Q_{\rm neu}$, thermal conductivity, and the boundary conditions at each end of the integration.

\subsection{Accretion heating and neutrino cooling}
\label{subsection: Accretion_heating}
\setcounter{equation}{0}
\renewcommand{\theequation}{B.1.\arabic{equation}}
\renewcommand{\thesubsection}{\thesection.\arabic{subsection}}

The HZ EOS \citep{1990A&A...229..117H, 1990A&A...227..431H} is adopted for the crustal region. The values for the mass number A, nuclear charge Z, electron chemical potential $\mu_{e}$, and the mass fraction of free neutrons $X_{n}$ are listed in Table~2 of  \citet{1990A&A...229..117H}. These quantities are essential, as the nuclear composition evolves in discrete steps with increasing density, corresponding to the formation of successive electron-capture layers.

As the material moves deeper into the crust, electron-capture reactions begin to occur at a constant pressure, depositing heat into the crust \citep{1990A&A...227..431H}. The heat deposited in each electron-capture layer per unit volume per unit time depends on the accretion rate $\dot{M}$.  The heat released in each capture layer is through smearing the heat deposited over whole shells as \citep{Hutchins:2022chj}
\begin{equation} \label{eq: nuc}
     Q_{\rm nuc} =
     \frac{\dot{M}\epsilon_{\rm{nuc}}}{\frac{4}{3}\pi(r^3_{i}-r^3_{i+1})} \,,
\end{equation}
where $\epsilon_{\rm nuc}$ is the heat deposited per nucleon by the relevant nuclear reaction, and $r_{i}$  is the radii at the $i^{\rm{th}}$ capture layer.  Almost the heat produced is released in the inner crust via pycnonuclear reactions \citep{1990A&A...229..117H, 1990A&A...227..431H, Fantina:2018yad}.

In the crustal region, the neutrino luminosity is dominated by the
electron-ion bremsstrahlung \citep{Brown:1999dk}.  The local energy loss due to
neutrino emission $Q_{\rm neu}$ is given by \citep{Haensel:1996rd} 
\begin{equation} \label{eq: neu_1}
     Q_{\rm neu} = 3.229 \times 10^{17} \rho_{12}
     T^{6}_{9}\frac{Z^2}{A}(1 - X_{n}) \, \rm erg \, s^{-1}\,cm^{-3} \,,
\end{equation}
where $\rho_{12} = \rho/10^{12}\, \rm g\,cm^{-3}$ and $T_{9} = T/10^9\, \rm K$.

\subsection{Thermal conductivity}
\label{subsection: Thermal_conductivity}
\setcounter{equation}{0}
\renewcommand{\theequation}{B.2.\arabic{equation}}
\renewcommand{\thesubsection}{\thesection.\arabic{subsection}}

The thermal conductivity in the crust helps determine the temperature profile
and heat flux, which is very important for the cooling and transport properties
of the NS \citep{Potekhin:2015qsa}. Using the results of
\citet{Yakovlev:2000jp}, the conductivity can be written as
\begin{equation} \label{eq: conductivity_1}
    \kappa = \frac{\pi^2}{3} \frac{k_{B}^2 T n_{e}}{m^{*}_{e}} \tau
    \,.
\end{equation}
Here, $n_{e}$ is the electron number density, $m^{*}_{e}$ is the effective
electron mass, and $\tau = {1}/{\nu}$ is the relaxation time where $\nu$ is the
scattering frequency of the different  mechanisms.  In the crustal region, the scattering frequency $\nu$ is determined as 
\begin{equation} \label{eq: tau_1}
   \nu = \nu_{\rm ep} + \nu_{\rm eQ} \,,
\end{equation}
where $\nu_{\rm ep}$ and $\nu_{\rm eQ}$ are the scattering frequencies for the
electron-phonon and electron-impurity, respectively. Note that we neglect the term of the
electron-electron scattering frequency since the strong degeneracy of the
electrons restricts the available phase space \citep{Brown:1999dk}.

 We follow the formalism of \citet{1980SvA....24..303Y}, and the electron-phonon
 scattering is temperature dependent as
\begin{equation} \label{eq: tau_2}
  \nu_{\rm ep} = \frac{12 e^2 k_{B} T}{\hbar^2 c} \,.
\end{equation}
As the density increases,  electron-impurity scattering becomes more important,
and the scattering frequency is written as
\begin{equation} \label{eq: tau_3}
  \nu_{\rm eQ} = \frac{4\pi Q_{\rm imp} e^4 n_{\rm ion}}{p^2_{F} v_{F}}
  \Lambda_{\rm imp} \,,
\end{equation}
where $p_{F}$ and $v_{F}$ are the momentum and velocity of electrons at the
Fermi surface, respectively. $\Lambda_{\rm imp} \simeq 1$ is the logarithmic Coulomb factor. The impurity parameter is defined by 
\begin{equation}
 Q_{\rm imp}  \equiv  n_{\rm ion}^{-1} \sum_{i} n_{i} (Z_{i} - \langle Z \rangle
 )^2 \,.
\end{equation}
Here, the sum $i$ is over all the different species of ions, with atomic number
$Z_{i}$ and mean atomic number $\langle Z \rangle$. The impurity parameter in
the accreted crust remains highly uncertain. Observational results of the
cooling NS transient MXB 1659$-$29 indicate that the impurity parameter was
constrained to $Q_{\rm imp} < 10 $ \citep{Brown:2009kw}.
However, \citet{Ootes:2019uvd} investigated the long-term temperature evolution of a transiently accreting NS and suggested that the impurity parameter can range from $30$ to $100$. In our work, we set the impurity parameter to be $Q_{\rm imp} =10 $.

\subsection{Boundary conditions }
\label{sec: BC_1}
\setcounter{equation}{0}
\renewcommand{\theequation}{B.3.\arabic{equation}}
\renewcommand{\thesubsection}{\thesection.\arabic{subsection}}

In this subsection, we describe the boundary conditions to solve  Eqs.~(\ref{eq: dFdr}) and (\ref{eq: dTdr}).  The bottom boundary is the interface between the NS core and crust.  We assume that the core is isothermal, maintaining
the same temperature as the bottom boundary.  The bottom boundary can be written
as 
\begin{equation}
 F_{\rm bottom} = - \frac{L_{\rm core}}{4 \pi R_{\rm bottom}} \,,
\end{equation}
where $R_{\rm bottom}$ is the raduis of NS core. The neutrino luminosity of the
core due to the modified Urca formula is \citep{Shapiro:1983du}
\begin{align}\label{eq: BC_1}
L_{\rm core} &  = 5.93 \times 10^{39} \, \rm erg \,s^{-1}\,
\left(\frac{\mathit{M}}{\mathit{M}_{\odot}} \right) \left(\frac{\rho_{\rm nuc}}{\rho}
\right)^{1/3} \mathit{T}^8_8 \exp\left(-\frac{\Delta}{\mathit{k}_{\rm B} \mathit{T}_{\rm bottom}} \right) \,,
\end{align}
where $\rho_{\rm nuc}$ is nuclear density. In this work, we only consider the
normal core, for which $\Delta = 0$. Importantly, at the surface of the NS crust, the boundary condition is set by the characteristic temperature at the base of the accretion column, and the characteristic temperature is dependent on the magnetic fields (see Fig. \ref{fig: accretion_column}).

\section{Temperature perturbation calculation}
\label{sec: Appendix_III}
\setcounter{equation}{0}
\renewcommand{\theequation}{C\arabic{equation}}

Before describing our model, in order to provide a rough estimate of the impact of magnetic fields on thermal conductivity, we first calculate the dimensionless magnetisation parameter \citep{1980SvA....24..425U}, which is given by \citep{Osborne:2019iph} 
\begin{equation} 
\omega \tau \equiv \frac{e B}{m_e^* c} \tau 
\approx 4 \times 10^{-5} B_{9} \,,
\end{equation}
where $B_{9} = B/10^9 \, \rm G$, $e$ is the electronic charge and  $m_e^*$ is the relativistic mass of the electron, such that $\omega \equiv e B / (m_e^* c)$ is the gyrofrequency of the electron, and $\tau$ their scattering timescale.
For the typical value of $B = 10^{13} \, \rm G$ considered in this work, we find $\omega \tau \thicksim 0.01$, confirming that the perturbative formalism remains applicable.

Following the methodology of \citet{Osborne:2019iph}, we assume a spherically symmetric, steady-state temperature profile for magnetized NSs undergoing super-Eddington column accretion. From energy conservation, the heat flux $\mathbf{F}$ is related to the net heat generation rate per unit volume $Q$ via
\begin{equation} 
\nabla \cdot {\bf F} = Q \,.
\end{equation}
Assuming that the magnetic-field-induced variation in the net heating rate depends only on temperature \citep{Pons:2007vf, Osborne:2019iph}, the perturbation equation can be written as
\begin{equation}\label{eq: AppendixB_1}
\nabla \cdot {\bf \delta F} 
= \frac{{\rm d} Q}{{\rm d} T} \delta T \,.
\end{equation}
Owing to the spherical symmetry of the background magnetized NS, we expand the perturbed quantities $\delta T$ and $\mathbf{\delta F}$ in tensor spherical harmonics \citep{Osborne:2019iph}:
\begin{align}
 \delta T (r\,, \theta \,, \phi)  & = \sum_{\ell m} \delta T_{\ell m}(r) Y_{\ell m}(\theta \,, \phi)  \label{eq: AppendixB_2} \,, \\
 \bf \delta F & = \sum_{\ell m} U_{\ell m} \hat{\bf r}Y_{\ell m}  + V_{\ell m} \nabla Y_{\ell m} \label{eq: AppendixB_3} \,, 
\end{align}
where $\hat{\bf r}$ is the radial unit vector. Substituting Equations~(\ref{eq: AppendixB_2}) and (\ref{eq: AppendixB_3}) into Equation~(\ref{eq: AppendixB_1}) yields the perturbed thermal structure equations:
\begin{align}
 \frac{{\rm d}U_{\ell m}}{{\rm d} r}  & =
  -\frac{2}{r}U_{\ell m} + \frac{\ell(\ell+1) }{r^2}V_{\ell m}
  + \frac{{\rm d} Q}{{\rm d} T} \delta T_{\ell m}
  \label{eq: AppendixB_4} \,, \\
 \frac{{\rm d}(\delta T_{\ell m})}{{\rm d} r} & = 
 - \frac{1}{\kappa}\frac{{\rm d} \kappa}{{\rm d} T} \frac{{\rm d} T}{{\rm d} r}\delta T_{\ell m}
  - \frac{U_{\ell m}}{\kappa}
  \label{eq: AppendixB_5} \,,
\end{align}
accompanied by the algebraic relation,
\begin{equation}
V_{\ell m} = - \kappa \delta T_{\ell m} + \kappa \omega^{*}r 
\frac{{\rm d} T}{{\rm d}r}\psi_{\ell m}\,,
\end{equation}
where $\omega^{*} \approx 0.02 B_{12}/T_{9}$, with $B_{12} = B/10^{12}\, \rm G$, and $T_{9} = T/10^{9}\, \rm K$. A detailed description of the toroidal magnetic field structure $\psi_{\ell m}$ can be found in Equation~(33) of \citet{Osborne:2019iph}. In this work, we consider only the quadrupole component, i.e., $\ell = m = 2$. 
For the background solution, i.e., the angle-averaged temperature and thermal flux, the accretion column calculation provides a boundary condition at the base of the column, as it is in thermal equilibrium with the deeper crustal layers. Lateral heat conduction efficiently distributes this energy across the crust, establishing an approximately uniform background temperature. We elaborate the boundary conditions at both the bottom and the surface of the NS crust in Appendix \ref{sec: BC_1}.  The temperature perturbations, capturing angular asymmetries induced by magnetically regulated heat transport, are computed separately using linear perturbation theory. This approach is justified as the magnetization parameter remains small, ensuring that the perturbations represent only a minor correction to the background. Using the column base temperature to set the spherically symmetric background thus provides a consistent starting point for computing the perturbations, while avoiding double counting of local column-induced asymmetries. For these perturbations, we adopt the standard boundary conditions of \citet{Ushomirsky:2000ax, Osborne:2019iph}, with zero temperature perturbation at both the bottom and the surface of the NS crust.

On the other hand, we need to specify a form for the toroidal magnetic field. We are interested only in the $l=m=2$ spherical harmonic.  Following the approach in \citet{Osborne:2019iph}, we adopt the form
\begin{equation}
\psi_{22}(r) = C [(r-R_{\rm IB})(r-R_{\rm OB})]^2 \,,
\end{equation}
where $C$ is a constant, ensuring that the magnetic field is non-zero only within the computational domain; $R_{\rm IB}$ denotes the radius at the bottom of the crust, and $R_{\rm OB} $ is the radius at the surface of the NS crust. Such a choice results in a magnetic field that
\begin{equation}
 {\bf B} = C \frac{1}{4} \sqrt{\frac{15}{2\pi}} [(r-R_{\rm IB})r-R_{\rm OB})]^2 (-\sin\theta \sin2\phi {\bf e_\theta} + \sin\theta \cos\theta \cos2\phi {\bf e_\phi} ) \,.
\end{equation}
The magnetic field strength $B$ is a function of position. We will report the field strength at the point $\theta = \pi/2$, $\phi = \pi/4$, and halfway between $r=R_{\rm IB}$ and $r= R_{\rm OB}$ when describing the strength of a given magnetic configuration.

\section{Crust deformation calculation}
\label{sec: Appendix_IV}
\setcounter{equation}{0}
\renewcommand{\theequation}{D\arabic{equation}}

When a star deviates from perfect sphericity, it develops multipole moments, defined as
\begin{equation}\label{eq: Multipole}
    Q_{\ell m} \equiv \int_0^R \delta \rho_{\ell m}(r) \, r^{\ell + 2} \, {{\rm d} r} \,,
\end{equation}
where $(l, m)$ denotes the harmonic mode of the density perturbation $\delta
\rho_{\ell m}(r)$, $R$ is the stellar radius.
The quadrupole moment with $(\ell, m) = (2, 2)$ contributes to the CGW signal. \citet{Bildsten:1998ey} pointed out that the spin-up torque from accretion can be balanced by the emission of quadrupolar gravitational radiation. Under this torque-balance condition, the CGW amplitude can be expressed as \citep{Ushomirsky:2000ax}
\begin{equation}\label{eq: strain}
   h_{0} = \frac{32 \pi}{5} \left(\frac{\pi}{3}\right)^{1/2}\frac{G}{c^4}
   \frac{Q_{22}\, f^2_{\rm GW}}{d} \,,
\end{equation}
where $G$ and $c$ are the speed of light and the gravitational constant,
respectively, $d$ is the distance of the source, and the GW frequency is $ f^2_{\rm GW} = 2 \nu$ for a NS rotating with spin frequency $\nu$ \citep{LIGOScientific:2019yhl, Reed:2021scb}. 
Following our previous work \citep{Li:2024dvx}, the perturbed Euler equation for NSs with
an elastic crust in the Cowling approximation is written as
\begin{equation}\label{eq: elastic_per}
0 = - \nabla_i \delta P - \delta \rho \nabla_i \Phi + \nabla^j t_{i j} \,,
\end{equation}
where  $\Phi$ is the gravitational potential, and the shear-stress tensor $t_{i
j}$ describes elastic forces as a first-order perturbation. In our calculation,
the shear-stress tensor $t_{i j}$ can be written as 
\begin{equation}
    t_{i j} = \mu \left( \nabla_i \xi_j + \nabla_j \xi_i - \frac{2}{3} g_{i j}
    \nabla_k \xi^k \right) \,,
\end{equation}
where $g_{i j}$ is the flat three-metric, and the shear modulus $\mu$ in the
crust is,
\begin{equation}\label{eq: mu}
\mu = 0.1194 \frac{n_{\rm ion} (Ze)^2}{a}\,,
\end{equation}
where $a=[3/(4\pi n_{\rm ion})]^{1/3}$ is the average ion spacing.

Employing tensor spherical harmonics leads to a convenient decomposition of the
shear-stress tensor \citep{Ushomirsky:2000ax, Haskell:2006sv, Li:2024dvx} as
\begin{align}\label{eq: shear-stress tensor}
t_{i j} =& t_{rr}Y_{\ell m}(\nabla_i r \nabla_j r -\frac{1}{2}e_{i j})+t_{r
\perp}(r)f_{i j} \nonumber \\
&+ t_{\Lambda}(r)(\Lambda_{i j}+\frac{1}{2}Y_{\ell m}e_{i j}) \,, 
\end{align}
where $t_{rr}$, $t_{r \perp}$, and $ t_{\Lambda}$ are functions of radial
coordinate $r$, $\beta = \sqrt{\ell (\ell + 1)}$ \citep{Ushomirsky:2000ax, Haskell:2006sv, Li:2024dvx},  
\begin{eqnarray}
e_{i j} & = & g_{i j}- \nabla_i r \nabla_j r  \,, \\
f_{i j}  & = & \frac{r}{\beta} (\nabla_i r \nabla_j Y_{\ell m} + \nabla_j r
\nabla_i Y_{\ell m}) \,,\\
\Lambda_{i j } & = & \left(\frac{r}{\beta} \right)^2 \nabla_i  \nabla_j Y_{\ell
m}+ \frac{1}{\beta}f_{i j} \,.
\end{eqnarray}
The displacement vector $\xi^i$ is associated with  polar perturbations, which
is given by, 
\begin{equation}
    \xi^i = \xi_{r}(r) \nabla^{i} r Y_{\ell m} + \frac{r}{\beta} \xi_{\perp}(r)
    \nabla^{i} Y_{\ell m} \,,
\end{equation}
where $\xi_{r}(r)$ and $\xi_{\perp}(r)$ are the radial and tangential components
of the displacement, respectively.  Again, the perturbed continuity equation is
determined as 
\begin{equation}\label{eq: Continuity}
    \delta \rho =  - \rho  \frac{\partial \xi_{r}}{\partial r} - \left( \frac{2
    \rho}{r} + \frac{\partial \rho}{\partial r}  \right) \xi_{r} + \beta \rho
    \frac{\xi_{\perp}}{r}  \,. 
\end{equation} 

Using four new variables \citep{Ushomirsky:2000ax, Li:2024dvx}
\begin{align}
    & z_{1} = \frac{\xi_{r}}{r} \label{eq: variables_1} \,, \\
    & z_{2} =\frac{t_{r r}}{P} - z_{1} {\frac{{\rm d} \ln P}{{\rm d} \ln r}}
    \label{eq: variables_2} \,, \\
    & z_{3} =\frac{\xi_{\perp}(r)}{\beta r}  \label{eq: variables_3}\,, \\
    & z_{4} = \frac{t_{r \perp}}{\beta P}  \label{eq: variables_4} \,, 
\end{align}
we obtain the following set of coupled ordinary differential equations \citep{Ushomirsky:2000ax, Li:2024dvx},
\begin{align}
\frac{{\rm d} z_{1}}{{\rm d} \ln r}& = - \left( 1 + 2
\frac{\alpha_{2}}{\alpha_{3}} - \frac{\alpha_{4}}{\alpha_{3}}   \right)  z_{1} +
\frac{1}{\alpha_{3}}z_{2}  \nonumber\\ 
 & \quad  + \ell (\ell + 1)\frac{\alpha_{2}}{\alpha_{3}}z_{3} +
 \frac{1}{\alpha_{3}}\Delta S   \,, \label{eq: bc_p1} \\
\frac{{\rm d} z_{2}}{{\rm d} \ln r}&=\left( U V - 4 V + 12 \Gamma
\frac{\alpha_{1}}{\alpha_{3}} - 4 \frac{\alpha_{1} \alpha_{4}}{\alpha_{3}}
\right) z_{1}  \nonumber\\ 
 & \quad   + \left( V - 4  \frac{\alpha_{1}}{\alpha_{3}} \right) z_{2} + \ell
 (\ell +1 ) \left(V-6\Gamma  \frac{\alpha_{1}}{\alpha_{3}} \right) z_{3}
 \nonumber\\ 
  & \quad  +  \ell (\ell +1)z_{4} - 4 \frac{\alpha_{1}}{\alpha_{3}}\Delta S \,, 
  \label{eq: bc_p2} \\
\frac{{\rm d} z_{3}}{{\rm d} \ln r}&=- z_{1} + \frac{1}{\alpha_{1}}z_{4}  \,,
\label{eq: bc_p3}\\
\frac{{\rm d} z_{4}}{{\rm d} \ln r}&= \left( V - 6 \Gamma
\frac{\alpha_{1}}{\alpha_{3}} +2
\frac{\alpha_{1}\alpha_{4}}{\alpha_{3}}\right)z_{1} -
\frac{\alpha_{2}}{\alpha_{3}} z_{2} \nonumber\\ 
 & \quad  + \frac{2}{\alpha_{3}}\Big\{ [2\ell(\ell+1) -1] \alpha_{1}\alpha_{2} +
 2[\ell (\ell+1) -1]\alpha^2_{1} \Big\}z_{3}  \nonumber\\ 
  & \quad + (V -3 )z_{4} + 2  \frac{\alpha_{1}}{\alpha_{3}} \Delta S  \,,
  \label{eq: bc_p4} 
\end{align}
where 
\begin{eqnarray}
U \equiv \frac{{\rm d} \ln g}{{\rm d} \ln r} +2 \,, \quad 
V \equiv \frac{\rho g r}{P} = - \frac{{\rm d} \ln P}{{\rm d} \ln r}  \,,  \quad
\quad \\ \nonumber
\alpha_{1}  \equiv  \frac{\mu}{P} \,, \quad 
\alpha_{2} \equiv  \Gamma - \frac{3}{2}\frac{\mu}{P} \,,  \quad
\alpha_{3}  \equiv \Gamma + \frac{4}{3} \frac{\mu}{P} \,,
\end{eqnarray}
and $g$ is the positive Newtonian gravitational acceleration.  For the  case
of $\delta T$ source term,  we have \mbox{$\Gamma = \left. ({\partial \ln P}/{\partial
\ln \rho})\right|_{T}$}, $\alpha_{4} = \left. ({\partial \ln P}/{\partial \ln T})
\right|_{\rho} ({{\rm d} \ln T}/{{\rm d} \ln r})$, and $\Delta S =
\left.({\partial \ln P}/{\partial \ln T})\right|_{\rho} ({\delta T}/{T})$.

To solve these equations, we need the boundary conditions at the interface between the core and crust, as well as the interface between the crust and the ocean, respectively.
At the interface between the core and crust, we have
\begin{equation}
z_{2} = Vz_{1} \,,  \quad z_{4} = 0 \,.
\end{equation}
However, at the interface between the crust and the ocean, the density may have
discontinuity, and we obtain the following boundary conditions, 
\begin{equation}
z_{2} = V \frac{\rho_{1}}{\rho_{\rm s}}z_{1} \,,  \quad z_{4} = 0 \,,
\end{equation}
where $(\rho_{\rm s} - \rho_{1})/\rho \approx 5 \times 10^{-5}$, $\rho_{\rm s} $
and $\rho_{1}$ are densities of elastic and liquid at the interface.
Finally, the perturbed multipole moments can be written as \citep{Ushomirsky:2000ax, Li:2024dvx}
\begin{eqnarray}\label{eq: Q_2} 
Q_{lm} =&-& \int^{r_{\rm top}}_{r_{\rm bot}}
  \frac{1}{V} \bigg \{ \ell(\ell+1)z_{4} - 2 \alpha_{1}\left( 2\frac{{\rm d}
  z_{1}}{{\rm d} \ln r} + \ell(\ell+1)z_{3} \right) \nonumber \\ 
	&+& \left(\ell+4 - U \right) \left(z_{2} - V z_{1} \right) \bigg \}
	r^{\ell+2}\rho \, {\rm d} r \,,
\end{eqnarray}
where $r_{\rm top}$ and $r_{\rm bot}$ correspond to the top and bottom of the crust, respectively. In this work, we calculate the quadrupole moment, $Q_{22}$, using the equation above. 
The quadrupole moments obtained from our calculations are $4.5 \times 10^{37}\,\rm g\, cm^2$ for $B = 10^{12}\, \rm G$, $8.6 \times 10^{37}\, \rm g\,cm^2$ for $B = 5 \times 10^{12} \, \rm G$, and $1.2 \times 10^{38} \, \rm g\,cm^2$ for $B = 10^{13}\,\rm G$, respectively. The stellar ellipticity is calculated from the mass quadrupole moment using $ \epsilon = \sqrt{\frac{8 \pi}{15}} \frac{Q_{22}}{I_{\rm zz}}$, where $I_{\rm zz} \simeq 10^{45}\, \rm g\, cm^2$ is the principal moment of inertia. Using this relation, we obtain ellipticities of $\epsilon = 3.8 \times 10^{-8}$ for $B = 10^{12} \, \rm G$, $\epsilon = 7.2 \times 10^{-8}$ for $B = 5 \times 10^{12} \, \rm G$, and $\epsilon = 1.0 \times 10^{-7}$ for $B = 10^{13} \, \rm G$. These numerical results indicate that the ellipticity increases with magnetic-field strength and lies in the range $\epsilon \sim 10^{-8}$ -- $10^{-7}$ for the parameter space relevant to ULXPs. The results can therefore be approximately summarized by the scaling relation $\epsilon  \sim  (1 - 2 ) \times 10^{-7}\, B_{12}$, which should be interpreted as an order-of-magnitude estimate inferred from the numerical calculations.


\bibliography{ULXPs_CGW.bib}{}
\bibliographystyle{aasjournalv7}

\end{document}